\documentclass[a4paper,11pt]{article}
\usepackage{pos}
\usepackage[normalem]{ulem}
\newcommand{\be}{\begin{equation}}
\newcommand{\ee}{\end{equation}}
\newcommand{\bea}{\begin{eqnarray}}
\newcommand{\eea}{\end{eqnarray}}

\newcommand{\MSbar}{{\overline{\rm MS}}}

\newcommand{\la}{\lambda}

\def\slashed{{/}\mskip-10.0mu}

\title{Perturbative Renormalization of Noether Currents in SYM with Improved Discretizations }

\author*[a]{Marios Costa}
\author[b]{Elias Antoniou}
\author[b]{Salomi Dimou}
\author[b]{Constantinos Costa}
\author[c]{Panos K. Chrysanthis}
\author[d]{Gregoris Spanoudes}
\author[d]{Haralambos Panagopoulos}

\affiliation[a]{Department of Mechanical Engineering and Material Science and Engineering, Cyprus University of Technology, Limassol, CY-3036, Cyprus}
\affiliation[b]{Rinnoco Ltd, Limassol, CY-3047, Cyprus}
\affiliation[c]{Department of Computer Science, University of Pittsburgh, Pittsburgh, PA-15260, USA}
\affiliation[d]{Department of Physics, University of Cyprus, Nicosia, CY-1678, Cyprus}

\emailAdd{marios.costa@cut.ac.cy}
\emailAdd{elias.a@rinnoco.com}
\emailAdd{salomi@rinnoco.com}
\emailAdd{costa.c@rinnoco.com}
\emailAdd{panos@cs.pitt.edu}
\emailAdd{spanoudes.gregoris@ucy.ac.cy}
\emailAdd{panagopoulos.haris@ucy.ac.cy}

\abstract{We investigate the perturbative renormalization of Noether currents associated with chiral symmetry and supersymmetry transformations in $\mathcal{N}=1$
 supersymmetric Yang–Mills theory, regularized on the lattice using improved discretizations. In particular, we study the chiral current and the supercurrent operator for a broad class of lattice actions: Wilson fermions with one-step stout-smeared links and a clover term, together with a family of Symanzik-improved gluon actions. Since the supercurrent mixes with both gauge-invariant and gauge-noninvariant operators, we compute the corresponding renormalization factors and mixing coefficients at one-loop order. We analyze the impact of improved gluon and gluino actions on this mixing pattern and investigate whether suitable choices of lattice parameters, such as the clover and stout coefficients, can reduce unwanted lattice artifacts at one loop. Our results provide perturbative input for the implementation of Ward identities and for matching lattice observables to their continuum counterparts, while also offering guidance for simulations aimed at restoring supersymmetry in the continuum limit.}

\FullConference{The 43rd International Symposium on Lattice Field Theory (LATTICE2026)\\
July 26 to August 1, 2026,\\
University of Maryland, College Park, USA\\}

\begin{document}
\maketitle

\section{Introduction}
\label{sec:introduction}

Supersymmetry (SUSY) provides an appealing extension of the Standard Model, addressing the hierarchy problem and offering viable dark-matter candidates~\cite{ParticleDataGroup:2026mpi}. A particularly useful framework for investigating its nonperturbative properties is $\mathcal{N}=1$ supersymmetric Yang--Mills (SYM) theory, which describes gluons and their fermionic superpartners, the gluinos, while retaining many features of more general supersymmetric gauge theories.

The lattice formulation of SUSY theories is challenging because discretization explicitly breaks supersymmetry at finite lattice spacing~\cite{Ali:2019agk,Bergner:2015adz,Feo:2002yi,Kaplan:2007zz,Catterall:2009it}. Its restoration in the continuum limit therefore requires appropriate fine-tuning, which can be studied through SUSY and chiral Ward identities~\cite{Curci:1986sm,Farchioni:2001wx,Feo:2003km,Taniguchi:1999fc,Ali:2018fbq}. This procedure is complicated by the mixing, under renormalization, of the supercurrent operator and chiral current with other operators allowed by the lattice symmetries.

In this work, we extend previous perturbative studies~\cite{Bergner:2022wnb} by performing high-precision one-loop calculations of the renormalization and mixing coefficients of the supercurrent and chiral current. We incorporate stout-smeared gauge links both in the gluino action and in the definition of the operators on the lattice, and consider three commonly used Symanzik-improved gluon actions. The resulting lattice renormalization and mixing coefficients are expressed in the $\MSbar$ scheme. Further details of the theoretical and computational framework can be found in Ref.~\cite{Costa:2026zwa}; here we focus on the new lattice results and their implications for future studies, including extensions to supersymmetric QCD.

\section{Computational Setup}
\label{sec:setup}
We employ the standard lattice discretization, with gluinos $\lambda(x) = \lambda^\alpha T^\alpha$ ($T^\alpha$: $SU(N_c)$ generators) defined on lattice sites and gluons represented by link variables,
\begin{equation}
U_\mu(x)=e^{igaT^\alpha u_\mu^\alpha(x+a\hat{\mu}/2)}.
\end{equation}
Since lattice discretization explicitly breaks supersymmetry and chiral symmetry, appropriate fine-tuning is required to recover these symmetries in the continuum limit. To improve the behavior at finite lattice spacing and suppress ultraviolet fluctuations, we employ stout-smeared links~\cite{Morningstar_2004, Horsley_2008},
\begin{equation}
\widetilde U_\mu(x)=e^{iQ_\mu(x)}U_\mu(x),
\end{equation}
where $Q_\mu(x)$ is constructed from the staples surrounding \(U_\mu(x)\) and depends on the stout parameter \(\omega\); its explicit form is given by:
\begin{equation}
  Q_\mu(x) = \frac{\omega}{2\,i} \left[V_\mu(x) U_\mu^\dagger(x) -  
  U_\mu(x)V_\mu^\dagger(x) -\frac{1}{3} \text{Tr} \,\Big(V_\mu(x)  
  U_\mu^\dagger(x) -  U_\mu(x)V_\mu^\dagger(x)\Big)\right] \, .
\label{Q_def}
\end{equation}  
Here, the ``stout coefficient'', $\omega$, is a tunable parameter, while $V_\mu(x)$ represents the sum over all staples associated with the link $U_\mu(x)$. The subscripts ``$A$'' and ``$O$'' on the stout coefficient, $\omega_A$, $\omega_O$, refer to its origin, with $A$ denoting the action and $O$ the operator.

The total lattice action, $S_{tot}^{LR}$, employed in our calculation is:
\begin{equation}
S^{LR}_{tot} = S^{LR}_{\rm SYM} + S_{\rm gf} + S_{\rm FP}  + S_{\rm meas.}\, ,
\label{ActionFull}
\end{equation}
where $S^{LR}_{\rm SYM}$ is shown in Eq.~\eqref{susylagrLattice}, while $S_{\rm FP}$, $S_{\rm gf}$, and $S_{\rm meas.}$ denote the discrete version of gauge-fixing, Faddeev--Popov, and measure contributions, as defined in Eqs.~\eqref{gaugefixing}, \eqref{ghosts}, and \eqref{measure}, respectively.

\subsection{Lattice improved actions and definitions of operators}

In our lattice formulation, gluinos are represented by clover-improved Wilson fermions in the adjoint representation; link variables present in covariant derivatives are stout-smeared, while those within the clover term are not. For the gauge fields we employ a class~\cite{Horsley_2008} of Symanzik improved gauge actions, involving Wilson loops with 4 and 6 links ($1\times1$ plaquettes and $1\times2$ rectangles, respectively). The Euclidean action ${\cal S}^{LR}_{\rm SYM}$ on the lattice then takes the form\footnote{Here, and in what follows, all traces are taken over color indices.}:
\begin{equation}
\begin{split}
{\cal S}^{LR}_{\rm SYM} \;=\;  \sum_{x} \Bigg\{ 
& \frac{2}{g^2} \Bigg[ 
   c_0 \sum_{\rm plaq} \, {\rm Re\,Tr}\!\left( 1 - U_{\rm plaq} \right) 
 + c_1 \sum_{\rm rect} \, {\rm Re\,Tr}\!\left( 1 - U_{\rm rect} \right) 
\Bigg] \\[6pt]
& + a^{4} \sum_{\mu} \Bigg[
   {\rm Tr}\!\left( \bar\lambda \gamma_{\mu} \Tilde{D}_{\mu} \lambda \right)
 - \frac{a r}{2} \, {\rm Tr}\!\left( \bar\lambda \Tilde{D}^{2} \lambda \right)
\Bigg] - a^{4}\sum_{\mu,\nu} \Bigg[
   \frac{c_{\rm SW}\,a}{4} \, 
   \bar\lambda^{\alpha} \sigma_{\mu \nu} \hat{F}_{\mu \nu}^{\alpha \beta} \lambda^{\beta}
\Bigg]
\Bigg\},
\end{split}
\label{susylagrLattice}
\end{equation}
where  $\sigma_{\mu\nu}=\frac{1}{2}[\gamma_\mu,\gamma_\nu]$ and $\hat{F}_{\mu \nu}^{\alpha \beta}$ in the adjoint representation is defined as:
\bea
\hat{{F}}_{\mu \nu}^{\alpha \beta} \! \! &=& \! \! \frac{1}{8}({Q}_{\mu \nu}^{\alpha \beta}-{Q}_{\nu \mu}^{\alpha \beta}),\\
{Q}_{\mu \nu}^{\alpha \beta} \! \! &=& \! \! 2{\rm{Tr}} \bigg( 
T^\alpha\, U_{x,x+\mu}U_{x+\mu,x+\mu+\nu}U_{x+\mu+\nu,x+\nu}U_{x+\nu,x}T^\beta\,U_{x,x+\nu}U_{x+\nu,x+\mu+\nu}U_{x+\mu+\nu,x+\mu}U_{x+\mu,x} \nonumber\\
&+& \! \! T^\alpha\, U_{x,x+\nu}U_{x+\nu,x+\nu-\mu}U_{x+\nu-\mu,x-\mu}U_{x-\mu,x} T^\beta\, U_{x,x-\mu}U_{x-\mu,x-\mu+\nu}U_{x-\mu+\nu,x+\nu}U_{x+\nu,x} \nonumber\\
& +& \! \! T^\alpha\, U_{x,x-\mu}U_{x-\mu,x-\mu-\nu}U_{x-\mu-\nu,x-\nu}U_{x-\nu,x} T^\beta\, U_{x,x-\nu}U_{x-\nu,x-\mu-\nu}U_{x-\mu-\nu,x-\mu}U_{x-\mu,x}\nonumber\\
& +& \! \! T^\alpha\, U_{x,x-\nu}U_{x-\nu,x-\nu+\mu}U_{x-\nu+\mu,x+\mu}U_{x+\mu,x} T^\beta\, U_{x,x+\mu}U_{x+\mu,x+\mu-\nu}U_{x+\mu-\nu,x-\nu}U_{x-\nu,x}\bigg),
\eea
and the plaquette ($U_{\rm plaq}$) and rectangle ($U_{\rm rect}$) loop variables are given by:
\begin{align}
U_{\rm plaq}(x;\mu,\nu) &= U_\mu(x)\, U_\nu(x+a\hat\mu)\, 
U_\mu^\dagger(x+a\hat\nu)\, U_\nu^\dagger(x), \\
U_{\rm rect}(x;\mu,\nu) &= U_\mu(x)\, U_\mu(x+a\hat\mu)\, 
U_\nu(x+2a\hat\mu)\, U_\mu^\dagger(x+a\hat\nu+a\hat\mu)\,
U_\mu^\dagger(x+a\hat\nu)\, U_\nu^\dagger(x),
\end{align}
The coefficients $\{c_0,c_1\}$ can, in principle, be chosen arbitrarily, subject to the normalization condition:
\begin{equation}
c_0 + 8c_1 = 1,
\label{norm}
\end{equation}
which ensures the correct classical continuum limit of the action. In our calculations, we employ the following sets of coefficients:
\begin{equation}
\{c_0,c_1\}
=
\left\{
\{1,0\},
\left\{\frac{5}{3},-\frac{1}{12}\right\},
\{3.648,-0.331\}
\right\},
\end{equation}
which correspond to the Wilson, tree-level Symanzik, and Iwasaki gluon actions, respectively.

The definitions of the covariant derivatives are as follows:
\bea
\! \! \! {\Tilde{\cal{D}}}_\mu\lambda(x) \! \! &\equiv& \! \! \frac{1}{2a} \Big[ \Tilde{U}_\mu (x) \lambda (x +  a \hat{\mu}) \Tilde{U}_\mu^\dagger (x) - \Tilde{U}_\mu^\dagger (x - a \hat{\mu}) \lambda(x - a \hat{\mu}) \Tilde{U}_\mu(x - a \hat{\mu}) \Big], \\
\! \! \! {\cal D}^2 \lambda(x) \! \! &\equiv& \! \! \frac{1}{a^2} \sum_\mu \Big[ \Tilde{U}_\mu (x)  \lambda (x + a \hat{\mu}) \Tilde{U}_\mu^\dagger (x)  - 2 \lambda(x) +  \Tilde{U}_\mu^\dagger (x - a \hat{\mu}) \lambda (x - a \hat{\mu}) \Tilde{U}_\mu(x - a \hat{\mu})\Big].
\eea

The remaining constituents of the action in Eq.~\eqref{ActionFull} are expressed as follows, starting with the gauge-fixing term $S_{\rm gf}$:
\begin{align}
    S_{\rm gf} = \frac{1}{2\xi} a^2 \sum_x \sum_\mu{\rm{Tr}} \left(u_\mu(x + a \hat{\mu}/2)-u_\mu(x - a \hat{\mu}/2)\right)^2,
    \label{gaugefixing}
\end{align}
where $\xi$ is the gauge parameter. The ghost term $S_{\rm FP}$, following the Fadeev-Popov procedure, admits the following form:
\bea
S_{\rm FP} &=& 2a^2 \sum_x\sum_\mu {\rm{Tr}} \{(\bar c(x + a \hat{\mu}) - \bar c(x))( c(x + a \hat{\mu}) - c(x) \nonumber \\
&&+ ig [u_\mu (x+ a \hat{\mu}/2),c(x)] + \frac{1}{2}ig [u_\mu (x+ a \hat{\mu}/2),c(x+a \hat{\mu}) -c(x)]\nonumber\\
&&-\frac{1}{12}g^2[u_\mu (x+ a \hat{\mu}/2),[u_\mu (x+ a \hat{\mu}/2),c(x+a \hat{\mu}) -c(x)]])\}+{\cal{O}}(g^3).
\label{ghosts}
\eea
Finally, the measure term $S_{\rm meas.}$ is given by:
\begin{equation}
    S_{\rm meas.} = \frac{g^2N_c}{12}a^2\sum_x \sum_\mu {\rm{Tr}}\left(u_\mu(x+ a \hat{\mu}/2)\right)^2 + {\cal{O}}(g^4).
    \label{measure}
\end{equation}

Turning our focus to the operators, following Noether’s procedure, the chiral current and supercurrent are given by~\cite{MARTIN_1998,gates2001superspacethousandlessonssupersymmetry}:
\begin{equation}
J_{5\mu} = \mathrm{Tr}\left(\bar \la \, \gamma_5 \, \gamma_\mu \, \la \right), \quad
S_\mu = -\frac{1}{2} \, \mathrm{Tr} \left( u_{\rho\sigma} \, [\gamma_\rho, \gamma_\sigma] \, \gamma_\mu \lambda \right).
\end{equation}

The chiral current, $J_{5\mu}$, renormalizes multiplicatively, while the supercurrent $S_\mu$, being a dimension-$7/2$ composite operator, can mix with operators of equal or lower dimension allowed by the symmetries of the employed action. The relevant operators are classified as follows:
\begin{itemize}
  \item \textbf{Class G: Gauge-invariant operators.} 
  \begin{equation}
      T_\mu = 2\, \mathrm{Tr} \left( u_{\mu\nu} \, \gamma_\nu \, \lambda \right).
  \end{equation}

  \item \textbf{Class A: BRST-exact operators}~\cite{Suzuki:2012pc}\textbf{.}
  \begin{equation}
    \mathcal{O}_{A1} = \frac{1}{\xi} \, \mathrm{Tr} \left( (\partial_\nu u_\nu)\, \gamma_\mu \lambda \right)
    - i g\, \mathrm{Tr} \left( [c, \bar{c}]\, \gamma_\mu \lambda \right).
  \end{equation}

  \item \textbf{Class B: Operators vanishing by the equations of motion.} 
  \begin{equation}
    \mathcal{O}_{B1} = \mathrm{Tr} \left( u_\mu\, \slashed{D} \lambda \right), \quad
    \mathcal{O}_{B2} = \mathrm{Tr} \left( \slashed{u}\, \gamma_\mu\, \slashed{D} \lambda \right).
  \end{equation}

  \item \textbf{Class C: Operators compatible with global symmetries but not in the above classes.} \\
  \begin{align}
    \mathcal{O}_{C1} &= \mathrm{Tr} \left( u_\mu\, \lambda \right), \ 
    \mathcal{O}_{C2} = \mathrm{Tr} \left( \slashed{u}\, \gamma_\mu\, \lambda \right), \ 
    \mathcal{O}_{C3} = \mathrm{Tr} \left( \slashed{u}\, \partial_\mu \lambda \right), \
    \mathcal{O}_{C4} = \mathrm{Tr} \left( (\partial_\mu \slashed{u})\, \lambda \right), \\
    \mathcal{O}_{C5} &= \mathrm{Tr} \left( (\partial_\nu u_\nu)\, \gamma_\mu \lambda \right), \
    \mathcal{O}_{C6} = \mathrm{Tr} \left( u_\nu\, \gamma_\mu\, \partial_\nu \lambda \right), \\
    \mathcal{O}_{C7} &= i g\, \mathrm{Tr} \left( [u_\rho, u_\sigma]\, [\gamma_\rho, \gamma_\sigma]\, \gamma_\mu \lambda \right), \
    \mathcal{O}_{C8} = i g\, \mathrm{Tr} \left( [u_\mu, u_\nu]\, \gamma_\nu \lambda \right).
  \end{align}
\end{itemize}

Thus, the renormalized chiral current $J_{5\mu}^{R}$ is related to the corresponding bare current $J_{5\mu}^{B}$ through the multiplicative renormalization factor $Z_J$:
\begin{equation}
    J_{5\mu}^R = Z_J J^B_{5\mu}, \label{ZJ}
\end{equation}
while the renormalized supercurrent $S_\mu^R$ is expressed as the following linear combination:
\begin{equation}
S_\mu^R = Z_{SS} S_\mu^B + Z_{ST} T_\mu^B + Z_{SA1} {\cal O}_{A1}^B
+ \sum_{i=1}^{2} Z_{SBi} {\cal O}_{Bi}^B
+ \sum_{i=1}^{8} Z_{SCi} {\cal O}_{Ci}^B,
\label{Zz}
\end{equation}
where the superscript $B$ ($R$) denotes bare (renormalized) operators, and $Z_{\cal O O'}$ are the corresponding
renormalization/mixing coefficients. The coefficients $Z_{\cal O O'}$ are  also labeled by the
regularization, $X$, and the renormalization scheme, $Y$, used,
and are therefore written as $Z^{X,Y}_{\cal O O'}$.

We performed perturbative one-loop computations of two-point Green's functions with external gluinos for the chiral current and external gluon, gluino for the supercurrent in arbitrary covariant gauge for both operators. The two-point Green's functions with external gluon ($u_\mu$) and gluino ($\lambda$) fields, we considered are:
\be
\langle \bar \la J_{5\mu} \la \rangle, \quad \langle u_\nu S_\mu \bar \la \rangle. \label{cc2ptgf}
\ee
Furthermore, we calculated the following three-point Green’s functions of the supercurrent involving external gluon ($u_\mu$), gluino ($\lambda$), and ghost ($c$) fields:
\begin{align}
\langle u_\nu(q_1)\,u_\rho(q_2)\,S_\mu\,\bar{\lambda}(q_3)\rangle, \qquad
\langle c(q_2)\,\bar{c}(q_1)\,S_\mu\,\bar{\lambda}(q_3)\rangle.
\label{3ptGFs}
\end{align} 
The relevant field and coupling constant renormalizations are defined as:
\be
u_\mu^R = \sqrt{Z_u}\, u_\mu^B, \quad \lambda^R = \sqrt{Z_\lambda}\, \lambda^B, \quad c^R = \sqrt{Z_c}\, c^B, \quad
g^R = Z_g \mu^{-\epsilon} g^B,\label{Zla}
\ee
where the expressions for $Z_u$, $Z_\lambda$, $Z_c$ and $Z_g$ in this lattice formulation and in the $\MSbar$ scheme, can be found in the dataset of Ref.~\cite{salomi_dimou_2026_18839281}. The dimensional regularization scale is $\mu$,  and $\epsilon$ is related to the spacetime dimension by $d=4-2\epsilon$; the quantity $\mu$ is connected to the standard $\MSbar$ renormalization scale, $\bar \mu$, via: $\mu = \bar\mu \, (4\pi/e^{\gamma_E})^{-1/2}$ ($\gamma_E$ is the Euler constant).

Table~\ref{tb:treelevels} presents the tree-level Green’s functions of all relevant operators involving external gluon, gluino, and ghost fields. These expressions are required to extract the $Z$-factors appearing in Eq.~\eqref{Zz}, since the corresponding structures naturally arise in the one-loop Green’s functions.
{\tiny{
\begin{table}[h!]
\centering
\begin{tabular}{|c | c | c | c|}
\hline
 & Tree-Level Two-Point & Tree-Level Three-Point & Tree-Level Three-Point \\
Operator  & $\langle u_\nu^{\alpha_1}(-q_1)\,{\cal O}_{i}\, \bar\lambda^{\alpha_2}(q_2) \rangle $ & $ \langle u_\nu^{\alpha_1}(-q_1)\, u_\rho^{\alpha_2}(-q_2)\,{\cal O}_{i}\, \bar\lambda^{\alpha_3}(q_3) \rangle $ & $ \langle c^{\alpha_3}(q_3) \,{\cal O}_{i}\,\bar c^{\alpha_2}(q_2) \bar\lambda^{\alpha_1}(q_1) \rangle $ \\[0.5ex]
\hline\hline 
$S_\mu$ & $-i (\slashed{q_1} \gamma_\nu - q_{1\nu}) \gamma_\mu$ & $g [\gamma_\nu,\gamma_\rho] \gamma_\mu / 2$ & $0$ \\[2ex]
\hline
$T_\mu$ & $i (q_{1\mu} \gamma_\nu - \slashed{q_1} \delta_{\mu \nu})$ & $-g (\delta_{\mu \nu} \gamma_\rho - \delta_{\mu \rho} \gamma_\nu)$ & $0$ \\[2ex]
\hline
${\cal O}_{A1}$ & $i q_{1\nu} \gamma_\mu / (2\alpha)$ & $0$ & $(g/2) \gamma_\mu$ \\[2ex]
\hline
${\cal O}_{B1}$ & $i \delta_{\mu \nu} \slashed{q_2} / 2$ & $-g (\delta_{\nu \mu} \gamma_\rho - \delta_{\rho \mu} \gamma_\nu) / 2$ & $0$ \\[2ex]
\hline
${\cal O}_{B2}$ & $i \gamma_\nu \gamma_\mu \slashed{q_2} / 2$ & $\, g (\gamma_\nu \gamma_\rho \gamma_\mu - \delta_{\rho\nu}\gamma_\mu +  \delta_{\mu\nu}\gamma_\rho  - \delta_{\mu\rho} \gamma_\nu) $ & $0$ \\[2ex]
\hline
${\cal O}_{C1}$ & $\delta_{\mu \nu} / 2$ & $0$ & $0$ \\[2ex]
\hline
${\cal O}_{C2}$ & $\gamma_\nu \gamma_\mu / 2$ & $0$ & $0$ \\[2ex]
\hline
${\cal O}_{C3}$ & $i \gamma_\nu q_{2\mu} / 2$ & $0$ & $0$ \\[2ex]
\hline
${\cal O}_{C4}$ & $i \gamma_\nu q_{1\mu} / 2$ & $0$ & $0$ \\[2ex]
\hline
${\cal O}_{C5}$ & $i \gamma_\mu q_{1\nu} / 2$ & $0$ & $0$ \\[2ex]
\hline
${\cal O}_{C6}$ & $i \gamma_\mu q_{2\nu} / 2$ & $0$ & $0$ \\[2ex]
\hline
${\cal O}_{C7}$ & $0$ & $-g [\gamma_\nu, \gamma_\rho] \gamma_\mu$ & $0$ \\[2ex]
\hline
${\cal O}_{C8}$ & $0$ & $-g (\delta_{\nu \mu} \gamma_\rho - \delta_{\rho \mu} \gamma_\nu) / 2$ & $0$ \\[2ex]
\hline
%${\cal O}_{C9}$ & $0$ & $0$ & $-(g/2) \gamma_\mu$ \\[2ex]
%\hline%\hline
\end{tabular}
\caption{Two-point and three-point tree-level Green's functions of $S_\mu$ and $T_\mu$ as well as of gauge noninvariant operators which may mix with $S_\mu$\,.
$\langle u_\nu^{\alpha_1}(-q_1)\,{\cal O}_{i}(x)\, \bar\lambda^{\alpha_2}(q_2) \rangle_{amp}^{tree}$ and $\langle u_\nu^{\alpha_1}(-q_1)\, u_\rho^{\alpha_2}(-q_2)\,{\cal O}_{i}(x)\, \bar\lambda^{\alpha_3}(q_3) \rangle_{amp}^{tree}$ are shown apart from an overall factor of $\delta^{\alpha_{1}\alpha_{2}} e^{i\,x\cdot(q_1+q_2)}$ and $f^{\alpha_{1}\alpha_{2}\alpha_{3}} e^{i\,x\cdot(q_1+q_2+q_3)}$, respectively. Similarly, $\langle c^{\alpha_3}(q_3) \,{\cal O}_{i}(x)\,\bar c^{\alpha_2}(q_2) \bar\lambda^{\alpha_1}(q_1) \rangle_{amp}^{tree}$ is shown apart from an overall factor of $f^{\alpha_{1}\alpha_{2}\alpha_{3}} e^{i\,x\cdot(q_1+q_2-q_3)}$.
\label{tb:treelevels}}
\end{table}
}}
%------------------------------------------------------------------
\section{Results for the renormalization of the chiral current and the supercurrent operators}
\label{sec:results}

\subsection{One-loop Renormalization of the chiral current operator}

\begin{figure}[ht!]
\centering
\includegraphics[scale=0.3]{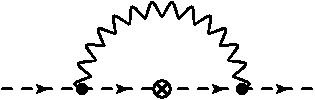}
\caption{One-loop Feynman diagram contributing to the two-point Green's function $\langle \lambda \, J_{5\mu} \, \bar \lambda \rangle$.  A wavy (dashed) line represents gluons (gluinos). A circled cross denotes the insertion of the operator.}
\label{fig2ptgJg}
\end{figure}

The one-loop correction to the Green's function in Eq.~\eqref{cc2ptgf} for the chiral current is determined by the diagram in Fig.~\ref{fig2ptgJg}. To determine the renormalization of the chiral current operator, we impose the following condition:
\begin{equation}
 \langle \lambda^R \, J_{5\mu}^R \,\bar{\lambda}^{R} \rangle_{amp} \Big{|}^\MSbar = Z_{\lambda}^{-1} Z_{J} \langle \lambda^B \, J_{5\mu}^B \,\bar{\lambda}^{B} \rangle_{amp}.
\label{GF2Bilinearconf}
\end{equation}

Once the $\MSbar$-renormalized and bare lattice Green’s functions have been calculated, the renormalization factor of the chiral current can be extracted directly. We obtain:
\begin{align}
Z_{J}^{LR,\,\overline{\mathrm{MS}}}
&= 1 + \frac{g^2 N_c}{16\pi^2}\,
\Big( {\rm{c}}^{J}_{00}\;
    + {\rm{c}}^{J}_{10}\,c_{\rm SW}\,r
    + {\rm{c}}^{J}_{20}\,c_{\rm SW}^2
    + {\rm{c}}^{J}_{01}\,\omega_A
    + {\rm{c}}^{J}_{11}\,c_{\rm SW}\,r\,\omega_A
    + {\rm{c}}^{J}_{02}\,\omega^2_A
\Big),
\label{ZJ5}
\end{align}
where the result for the one-loop Green's function for the chiral current, as well as the values for every numerical coefficient shown in Eq.~\eqref{ZJ5} can be found in Ref.~\cite{Costa:2026zwa} or in the dataset of Ref.~\cite{antoniou_2026_20268143}.

\subsection{One-loop mixing coefficients for the supercurrent operator}

\begin{figure}[ht!]
\centering
\includegraphics[scale=0.3]{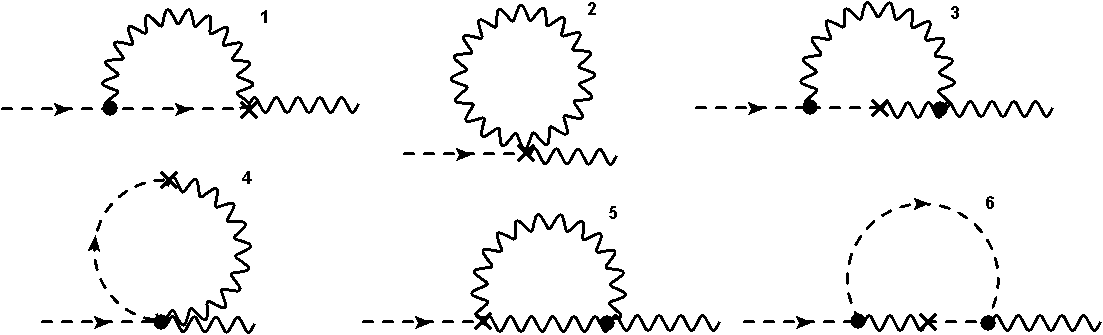}
\caption{One-loop Feynman diagrams contributing to the two-point Green's function $\langle u_\nu S_\mu \bar \lambda  \rangle$.  A wavy (dashed) line represents gluons (gluinos). A cross denotes the insertion of $S_\mu$. Diagrams 2, 4 do not appear in dimensional regularization; they do however show up in the lattice formulation.}
\label{fig2pt}
\end{figure}

At one loop, as an example, we present the Feynman diagrams associated with the two-point Green’s function of the supercurrent, shown in Fig.~\ref{fig2pt}. The one-loop Feynman diagrams contributing to the three-point Green’s function are shown in Figs.~2 and~3 of Ref.~\cite{PhysRevD.106.034502}.

To extract all the renormalization and mixing coefficients in Eq.~\eqref{Zz}, it is sufficient to evaluate the two-point Green’s function of the supercurrent operator for the following three choices of external momenta: $q_2=0$, $q_1=0$, and $q_2=-q_1$. Furthermore, we make a single choice of external momenta for the three-point Green’s functions of the supercurrent operator, which is sufficient in order to avoid the considerably more complicated expressions arising for generic external momenta. Results for all Green’s functions involving \(S_\mu\), together with the corresponding mixing coefficients, will be presented in a forthcoming publication; see Ref.~\cite{Costa:2026forthcoming}.

The strategy and the order in which the calculated Green’s functions are employed are crucial for determining the mixing coefficients unambiguously.

$\bullet\,$ We begin with the three-point Green’s function involving external gluino, ghost, and antighost fields, from which we determine the mixing coefficient associated with ${\cal O}_{A1}$:
\bea 
&& \hspace{-2cm} \langle c^R \,{S}^R_{\mu}\, \bar c^R \,\bar\lambda^R \rangle _{amp}\Big|^\MSbar = Z_\lambda^{-1/2} Z_c^{-1} Z_{SS}\langle c^B \,{S}^B_{\mu}\, \bar c^B \,\bar\lambda^B \rangle _{amp} + Z_{SA1} \langle c^B \, {\cal O}^B_{A1} \, \bar c^B\, \bar\lambda^B \rangle _{amp}^{tree} \,
+ {\cal O}(g^4). \label{3ptGFexprS}
\eea
We found:
\begin{equation}
Z_{SA1}^{LR,\MSbar}=0,
\end{equation}
demonstrating that the supercurrent does not mix with ${\cal O}_{A1}$ at one loop.

$\bullet\,$ We next consider the two-point Green’s function and impose the following renormalization condition at one-loop order for each of the three choices of external momenta:
\bea 
\langle u_\nu^R \,{S}^R_{\mu}\, \bar\lambda^R \rangle _{amp}\Big|^\MSbar &=& Z_\la^{-1/2} \,Z_u^{-1/2} Z_{SS} \langle u_\nu^B \,{S}^B_{\mu}\,\bar\lambda^B \rangle _{amp} + Z_{ST}  \langle u_\nu^B \, {T_\mu}^B\, \bar\lambda^B \rangle _{amp}^{tree} + Z_{SA1} \langle  u_\nu^B \, {\cal O}^B_{A1} \, \bar\lambda^B \rangle _{amp}^{tree}\nonumber\\ 
&+& \sum_{i=1}^{2} Z_{SBi} \langle  u_\nu^B \, {\cal O}^B_{Bi} \, \bar\lambda^B \rangle _{amp}^{tree} + \sum_{i=1}^{6} Z_{SCi} \langle u_\nu^B \,{\cal O}^B_{Ci} \, \bar\lambda^B \rangle _{amp}^{tree} + {\cal O}(g^4).
\label{2ptGFexprS}
\eea
%------------------------------------------------------------------
The choice $q_2=0$ allows us to determine the $Z$-factors for the operators $S_\mu$, $T_\mu$, $\mathcal{O}_{C4}$, and $\mathcal{O}_{C5}$:
\bea
\label{Zss}
%Z_{SS}^{LR,\MSbar} &=& 1 + \frac{g^2}{16\pi^2}\Bigg[ \frac{-9.86960}{N_c} + N_c(-2.3170 + 14.49751c_{\rm SW}^2 - 1.23662c_{\rm SW}\,r) \Bigg]\\
%Z_{SS}^{LR,\MSbar} = 1 &+& \frac{g^2}{16\pi^2\,N_c} \Bigg[-9.86960 + \left(315.8273 - 1421.22308\,\omega_O \right) \omega_O 
%\nonumber \\
%&& + N_c^2 \Big(-2.3169 - 369.8418\,\omega_A^2 + c_{\rm SW}^2 \left(14.49751 - 4.71777\,\omega_O \right) 
%\nonumber \\
Z_{SS}^{LR,\MSbar} &=& 1 + \frac{g^2}{16\pi^2} \Bigg[\frac{1}{N_c}\left(c^{SS}_{0000} + c^{SS}_{0001} \, \omega_O + c^{SS}_{0002} \, \omega_O^2 \right) + N_c \Big(c^{SS}_{1000} + c^{SS}_{1001} \, \omega_O
\nonumber \\
 && \hspace{-2cm} +  c^{SS}_{1002} \, \omega_O^2 + c^{SS}_{1010} \, \omega_A + c^{SS}_{1011} \, \omega_A \omega_O + c^{SS}_{1020} \, \omega_A^2 + c^{SS}_{1100} \, r c_{\rm SW} + c^{SS}_{1101} \, r c_{\rm SW} \omega_O
\nonumber \\
&& \hspace{-2cm}  + c^{SS}_{1110} \, r c_{\rm SW} \omega_A + c^{SS}_{1111} \, r c_{\rm SW} \omega_A \omega_O + c^{SS}_{1200} \, c_{\rm SW}^2 + c^{SS}_{1201} \, c_{\rm SW} \omega_O \Big) \Bigg], \\
\label{Zst}
Z_{ST}^{LR,\MSbar} &=& \frac{g^2}{16\pi^2} 3N_c,
\label{ZSA1Cs} \quad
Z_{SC4}^{LR,\MSbar} = Z_{SC5}^{LR,\MSbar}  = 0. 
\eea
Then, by imposing the renormalization condition at \(q_1=0\), the renormalization factors associated with the operators \(\mathcal{O}_{B1}\), \(\mathcal{O}_{B2}\), \(\mathcal{O}_{C3}\), and \(\mathcal{O}_{C6}\) can be extracted and are given by:
\bea
%Z_{SB1}^{LR,\MSbar} &=& \frac{g^2}{16\pi^2} N_c \Bigg[-0.38395 - 4.57771 \, \omega_O- \log\left(a^2 \bar \mu^2\right)\Bigg]\label{ZSB1}\\
%Z_{SB2}^{LR,\MSbar} &=&  \frac{g^2}{16\pi^2} N_c \Bigg[0.80802 -2.28886\, \omega_O -\frac{1}{2} \log\left(a^2 \bar \mu^2\right)\Bigg]\label{ZSB2}\\
Z_{SB1}^{LR,\MSbar} &=& \frac{g^2}{16\pi^2} N_c \Bigg[c^{SB1}_{0} + c^{SB1}_{1} \, \omega_O- \log\left(a^2 \bar \mu^2\right)\Bigg]\label{ZSB1}\\
\hspace{-2cm} Z_{SB2}^{LR,\MSbar} &=& \frac{g^2}{16\pi^2} N_c \Bigg[c^{SB2}_{0} + c^{SB2}_{1} \, \omega_O -\frac{1}{2} \log\left(a^2 \bar \mu^2\right)\Bigg],\label{ZSB2} \quad
Z_{SC3}^{LR,\MSbar} = Z_{SC6}^{LR,\MSbar}  = 0.
\eea
Finally, the case $q_2=-q_1$ provides an independent consistency check of the mixing coefficients determined above.

The complete expressions for the first row of the mixing matrix, for each gluon action, as well as the values for all numerical coefficients $c_i^X$ can be found in Ref.~\cite{antoniou_2026_20268143}. Note that we can choose the values of the clover parameter $c_{\rm SW}$ and the stout coefficients $\omega_A$ and $\omega_O$ so as to eliminate the finite contributions to these renormalization and mixing coefficients. However, this cancellation is valid only at one-loop order.

The operators $\mathcal{O}_{C1}$ and $\mathcal{O}_{C2}$ have lower dimension than the supercurrent and therefore, they can only mix with the supercurrent through terms containing inverse powers of the lattice spacing. Since such terms do not appear in the Green’s functions, these mixing are absent: 
\be
Z_{SC1}^{LR,\MSbar} = Z_{SC2}^{LR,\MSbar}  = 0
\ee

%\begin{figure}[ht!]
%\centering
%\includegraphics[scale=0.2]{GFguuLatt.png}\\
%\includegraphics[scale=0.2]{GFguuLatt2.png}
%\caption{One-loop Feynman diagrams contributing to the three-point Green's function $\langle u_\nu u_\rho S_\mu\bar \lambda  \rangle$.  A wavy (dashed) line represents gluons (gluinos). Diagrams 1, 2, 3, 4, 5, 6, 11, and 13 do not appear in dimensional regularization but they contribute in the lattice regularization. A cross denotes the insertion of the operator. A mirror version (under exchange of the two external gluons) of diagrams 3, 4, 5, 6, 8, 10, 14, 15 and 16 must also be included.}
%\label{fig3ptguu}
%\end{figure}

%\begin{figure}[ht!]
%\centering
%\includegraphics[scale=0.2]{GFGCCLatt.png}
%\caption{One-loop Feynman diagrams contributing to the three-point Green's function $\langle c \, S_\mu \, \bar c \, \bar \lambda \rangle$.  A wavy (dashed) line represents gluons (gluinos). A cross denotes the insertion of the operator. The ``double dashed'' line is the ghost field. Diagrams 1 and 2 do not appear in dimensional regularization; they do however show up in the lattice formulation.}
%\label{fig3ptgCC}
%\end{figure}

$\bullet\,$Lastly, the three-point Green’s function with two external gluons and one gluino is used to determine the mixing coefficients with $\mathcal{O}_{C7}$ and $\mathcal{O}_{C8}$. The external momenta are chosen such that $q_2=0$ and $q_3=-q_1$. The following renormalization condition is employed to extract the corresponding mixing coefficients:
\bea
&& \hspace{-2cm} \langle u_\nu^R u_\rho^R \,{S}^R_{\mu}\, \bar\lambda^R \rangle _{amp} \Big|^\MSbar = Z_\la^{-1/2} \,Z_u^{-1}\, Z_{SS} \langle u_\nu^B u_\rho^B \,{S}^B_{\mu}\,\bar\lambda^B \rangle _{amp} + Z_{ST} \langle u_\nu^B u_\rho^B \, {T_\mu}^B\, \bar\lambda^B \rangle _{amp}^{tree} \nonumber\\
&& \hspace{-2cm} +  Z_{SB1}\langle u_\nu^B u_\rho^B \, {\cal O}^B_{B1}\, \bar\lambda^B \rangle _{amp}^{tree} + Z_{SB2}\langle u_\nu^B u_\rho^B \, {\cal O}^B_{B2}\, \bar\lambda^B \rangle _{amp}^{tree} + \sum_{i=7}^{8} Z_{SCi}\langle u_\nu^B u_\rho^B \, {\cal O}^B_{Ci}\, \bar\lambda^B \rangle _{amp}^{tree} + {\cal O}(g^4).
\label{3ptGFexpr1S}
\eea
At one-loop order, we find that both mixing coefficients vanish:
\begin{equation}
   Z_{SC7}^{LR,\MSbar} = Z_{SC8}^{LR,\MSbar} = 0.
\end{equation}

All vanishing mixing coefficients remain zero for every gluon action considered, even in the presence of stout-smeared links. This constitutes a highly nontrivial check, given the number of free parameters involved in the lattice discretization: \(r=\pm1\), \(c_{\rm SW}\), \(\omega_A\), and \(\omega_O\).

\section{Summary and Future Plans}
\label{sec:conclusions}

In this work, we have determined the one-loop renormalization factor of the chiral current $J_{5\mu}$ and the one-loop mixing coefficients of the supercurrent $S_\mu$ in $\mathcal{N}=1$ supersymmetric Yang--Mills theory. The calculations were performed using gauge links subjected to one step of stout smearing and three commonly used Symanzik-improved gluon actions. The renormalization and mixing coefficients were extracted from the relevant two- and three-point Green’s functions.

As a natural extension of this work, we plan to investigate the renormalization of the gluino-glue operator within the same stout-smeared, Symanzik-improved lattice formulation.

%------------------------------------------------------------------
\section*{Acknowledgements}
 The projects EXCELLENCE/0524/0208 and EXCELLENCE/0524/0001 are implemented under the programme of social cohesion ``THALIA 2021-2027'' co-funded by the European Union, through Cyprus Research and Innovation Foundation. 
\bibliographystyle{JHEP}
\bibliography{references}

\end{document}